\documentclass{LT23auth}
\usepackage{graphicx}

\begin{document}

\begin{frontmatter}

\title{de Haas - van Alphen measurements on CeRhIn$_{\mathbf{5}}$ 
under pressure}

\author[NHMFL]{Donavan Hall \thanksref{thank1}},
\author[NHMFL]{S.~W.~Tozer},
\author[NHMFL]{E.~C.~Palm},
\author[NHMFL]{T.~P.~Murphy},
\author[LSU]{R.~G.~Goodrich},
\author[LANL]{J.~L.~Sarrao}

\address[NHMFL]{1800 E. Paul Dirac Dr. Tallahassee, Florida 32311, USA}

\address[LSU]{Louisiana State University, Department of Physics, Baton 
Rouge, Louisiana 70803, USA}

\address[LANL]{Los Alamos National Laboratory, MST-10, Los Alamos, New 
Mexico 87545, USA}

\thanks[thank1]{ E-mail:hall@magnet.fsu.edu}

\begin{abstract}
Measurements of the de Haas - van Alphen effect have been carried out 
on the heavy fermion anti-ferromagnet CeRhIn$_{5}$ at temperatures 
between 25 mK and 500 mK under pressure.  We present some preliminary results of our 
measurements to track the evolution of the Fermi surface as the 
pressure induced superconducting transition is approached.
\end{abstract}

%
%
\end{frontmatter}

\section{Introduction}

The dimensionality of the 115 materials, CeRhIn$_5$, CeIrIn$_5$, and
CeCoIn$_5$, appears to be related to their superconducting transition
temperature.  The material with the highest T$_{c}$, CeCoIn$_{5}$, has
the most 2D-like Fermi surface (FS) of the three.  \cite{Hallco}
CeRhIn$_{5}$ has a high T$_{c}$ (\ensuremath{\sim}2.1 K), but only
under a pressure of \ensuremath{\sim}16 kbar.  At ambient pressures,
CeRhIn$_{5}$ is an anti-ferromagnet.  The FS of CeRhIn$_{5}$ was the
subject of one of our recent publications.\cite{Hall2001} In order to
confirm the link between the superconducting state and FS
dimensionality, the FS as a function of pressure in CeRhIn$_{5}$
should be measured.  If the FS becomes more 2D-like as the critical
pressure is approached, then this will be evidence for making a
connection.

In these materials it seems that superconductivity does not appear 
until the overlap between the \textit{f} electron wavefunctions is 
sufficient to allow band-like behavior.  Measurements of the FS as a 
function of pressure should show this increasing overlap as a change 
in topography.  Here we present measurements up to 7.9 kbar, about 
half the critical pressure for CeRhIn$_5$.

\section{Results}

We have designed and built 
small pressure cells, capable of running in a dilution refrigerator 
and in a rotator. Measuring torque inside a pressure cell is 
impossible, so we have made small compensated pickup coils 
which fit into the cell. Each coil has four to five thousand 
turns. The filling factor approaches unity because we are able 
to situate the coil along with the sample inside the cell. A small 
coil is wound on the exterior of the cell to provide an 
ac modulation of the applied field.

We have measured the FS of CeRhIn$_{5}$ under several pressures. 
At each pressure we measure FS frequencies and their amplitude 
dependence as a function of temperature. From this we can extract 
information about how the effective mass of the quasiparticles 
is changing as the pressure is increased. The figures show the 
Fourier spectra of CeRhIn$_{5}$ under \ensuremath{\sim}7.9 kbar. The crystal 
was oriented so that the a-b axis plane is perpendicular to the 
applied field.

\begin{figure}[tbp]
\begin{center}\leavevmode
	\includegraphics[width=0.8\linewidth]{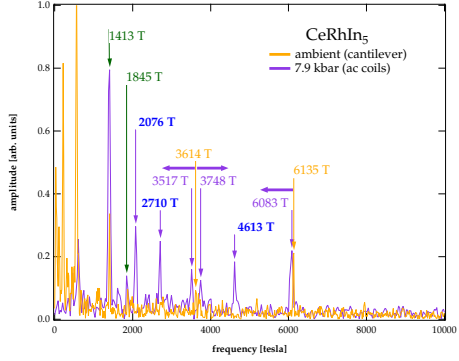}
	\caption{The higher frequency spectrum for the sample under 
	pressure shows
	the increase of amplitude of three peaks; an effect that can be 
	attributed to improved sample/coil coupling at higher pressures.}
\label{highfft}\end{center}
\end{figure}

\begin{figure}[tbp]
\begin{center}\leavevmode
	\includegraphics[width=0.8\linewidth]{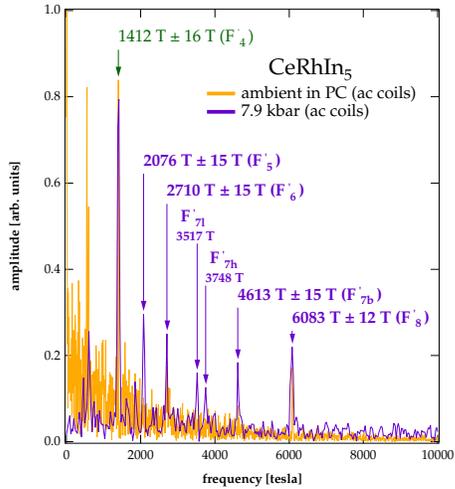}
	\caption{Comparison of the Fourier spectra of CeRhIn$_{5}$ at 
	\ensuremath{\sim}7.9 kbar and at ambient pressures (measured in the 
	pressure cell prior to pressurization) 
	reveals little that is suggestive of change.}
\label{lowfft}\end{center}
\end{figure}

%
%

\section{Discussion}

We show the 7.9 kbar data compared with two sets of data taken at ambient
pressure.  In Fig.  \ref{highfft} the FS at 7.9 kbar is compared with
the ambient data taken with a torque cantilever (the same data
reported in \cite{Hall2001}).  Because the modulation field for the ac
measurements (in the pressure cell) was so small, the lowest
frequencies can be ignored.  Notice that the 1411 T (F$^{'}_{4}$, the
designation given in Ref.  \cite{Hall2001}) and 1845 T peaks are
reproduced exactly in the ambient and the pressure data sets.  The
1845 T peak was not included in Ref.  \cite{Hall2001} because of its
small amplitude in ambient pressure torque measurements.

The 3600 T (F$^{'}_{7}$) 
and 6120 T (F$^{'}_{8}$) peaks are present in both data sets; however, 
the F$^{'}_{7}$ appears to have split and the F$^{'}_{8}$ appears to have 
shifted down in frequency.  Such changes 
could be explained as slight differences of sample alignment with 
respect to the applied field between the torque measurement and the 
pressure cell measurement. 
Three other frequencies, 2076 T, 2710 T, and 4613 T, emerge in the pressure 
data which are close to to some reported in Ref. \cite{Hall2001} to be 
observed only at the lowest temperatures (25 mK).

All but the first of these frequencies are seen also in ambient
pressure data taken with the sample in the pressure cell prior to
pressurization as shown in Fig.  \ref{lowfft}.  Thus, assuming the
differences in frequency between the torque measurements and pressure
cell measurements are due to differences in alignment, we can make
frequency assignments that follow Ref.  \cite{Hall2001} (also shown in
Fig.  \ref{lowfft}).  The relative increase in amplitude with
increasing pressure of these three peaks could be a result of the increase of the
coupling factor between the sample and the coil as the two are
compressed together.

The lack of any clear differences in the FS up to 7.9 kbar suggests
that if the FS changes, then such change is not a linear function of
pressure.  Nor is there a compelling reason to think that it should
be a linear function.  Possibly, at some pressure closer the the critical pressure, the
transition to \textit{f} electron itinerate behavior will take place 
leading to more noticable changes in the FS.

\section{Conclusions}

The FS of CeRhIn$_5$ appears to remain topographically stable under
the application of pressure up to 7.9 kbar.  Additional measurements
which approach the critical pressure ($\sim$16 kbar) are of
prime importance.
  
%
%
\begin{ack}
This work was performed 
at the National High Magnetic Field Laboratory, which is supported 
by NSF Cooperative Agreement No. DMR-9527035 and by the State 
of Florida. Work at Los Alamos was performed under the auspices 
of the U. S. Dept. of Energy.
\end{ack}

%
%

\end{document}